# Self-assembled structure of dendronized CdS nanoparticles

Hiroshi Nakajima[1,*], Daichi Matsuki[1], Yumi Fukunaga[2], Takaaki Toriyama[2], Koji Shigematsu[2], Masaki Matsubara[3], Kiyoshi Kanie[4], Atsushi Muramatsu[4], and Yasukazu Murakami[1, 2]

[1] *Department of Applied Quantum Physics and Nuclear Engineering, Kyushu University, Fukuoka 819-0395, Japan*

[2] *The Ultramicroscopy Research Center, Kyushu University, Fukuoka, 819-0395, Japan*

[3] *Department of Materials and Environmental Engineering, National Institute of Technology, Sendai College, Natori 981-1239, Japan*

[4] *Institute of Multidisciplinary Research for Advanced Materials, Tohoku University, Katahira 2-1-1, Sendai 980-8577, Japan*

*To whom correspondence should be addressed. E-mail: nakajima@nucl.kyushu-u.ac.jp

Self-assembled dendronized CdS nanoparticles have been attracting considerable attention because of their photoluminescence properties depending on annealing treatments. In this study, their annealing-induced self-assembled structure was investigated via scanning transmission electron microscopy (STEM); thin foil specimens of self-assembled dendronized CdS nanoparticles were prepared by ultramicrotomy and the STEM images revealed their ordered structure and the effect of the annealing treatment. In addition, a structural order belonging to the $P2_13$ space group was identified via an autocorrelation analysis. The results indicated that this structural order could be achieved only over a few tens of nanometers.

**Introduction**

Self-assembled two and three-dimensional arrays of organic and inorganic nanoparticles have attracted considerable attention because of their intriguing material functionalities [1–6] including photoluminescence, magnetic viscosity, catalytic activity, and optical properties. One useful method for controlling self-assembly is coating the nanoparticles with organic liquid crystals [7,8]. Kanie et al. fabricated organic–inorganic hybrid dendrimers, composed of gold nanoparticles (core) and organic dendrons (shell), and demonstrated the formation of a simple cubic structure by heating [9,10]. More recently, the method for the hybrid dendrimers has been applied to the fabrication of dendronized CdS nanoparticles, as shown in the schematic in Fig. 1(a) [11]; the CdS nanoparticles were modified by using a two-layered shell made of aliphatic thiols (inner shell) and liquid crystal aromatic dendrons (outer shell).

The self-assembly of CdS nanoparticles could be a promising route to achieving significant optical properties because of their photoluminescence given by quantum dots. Interestingly, the photoluminescence of dendronized CdS (hereafter, CdS-Dend) nanoparticles can be manipulated by post-dendronization annealing. Matsubara et al. [11] observed that the as-dendronized state (i.e., without annealing) yields a disordered configuration of CdS-Dend nanoparticles and exhibits significant photoluminescence that, however, is suppressed by post-dendronization annealing, which also seems to generate a structural order in the nanoparticles. Based on small-angle X-ray scattering measurements, they identified this structural order as belonging to the $P2_13$ space group (Fig. 1(b)). The photoluminescence suppression was explained by the non-radiative energy transfer via dendron π-stacking in the ordered phase. Such switchable photoluminescence is promising for applications in, e.g., photovoltaic cells, light-emitting diodes, and thermal sensors.

However, for future device applications, we need to further understand such a self-assembled structure. Annealing induces a structural order in CdS-Dend nanoparticles, but the spatial extent of this ordered phase is unclear. Although direct imaging of their self-assembled structure would provide useful information about the mechanism involved, similar studies are lacking. The crystallinity of individual CdS nanoparticles, which could affect the photoluminescence performance, is another issue. Electron microscopy could effectively solve these problems because it allows comprehensive studies of micro- and nanostructures.

This study aims to characterize the complex self-assembled structure of CdS-Dend nanoparticles by using scanning transmission electron microscopy. We focused on further understanding the spatial extent of the $P2_13$ structure obtained by annealing. For this purpose, we prepared via ultramicrotomy thin foil specimens of self-assembled nanoparticles. The microscopy images of the ordered phase were compared with those of the disordered one and analyzed on the basis of an autocorrelation function.

**Methods**

CdS-Dend nanoparticles were synthesized using a method described in [11]. They were dispersed in tetrahydrofuran and a droplet of the resulting solution was placed on a grid coated with an amorphous carbon film. To analyze the self-assembled structure, disordered and ordered specimens were prepared; the disordered specimen represented the case without post-dendronization annealing, while the ordered specimen was attained by annealing the CdS-Dend nanoparticles, dispersed in $CHCl_3$ on a hydrophobic quartz glass plate, at 423 K for 20 h under Ar atmosphere. In both the specimens, the CdS-Dend nanoparticles were condensed as a film and embedded in a resin rod, as shown in Fig. 2(a). After trimming into a shape favorable for ultramicrotomy, the rod was sliced into ~90 nm thick sections by using an ultramicrotome (UC6, Leica Co., Ltd.) (Fig. 2(b)); the thin sections were placed on copper grids with amorphous carbon meshes, as shown in Fig. 2(c).

High-angle annular dark-field scanning transmission electron microscopy (HAADF-STEM) images were obtained on a transmission electron microscope equipped with a spherical aberration corrector (JEM-ARM200CF, JEOL Co., Ltd.); the acceleration voltage, probe semi-angle, and current were 120 kV, 19 mrad, and 15 pA, respectively, and the angular detection range of the HAADF detector for scattered electrons was 50–130 mrad.

**Results and discussion**

The structure of individual CdS nanoparticles was first investigated. Figure 1(c) shows a HAADF-STEM image of five CdS-Dend nanoparticles. Although the dendrons (shells) were difficult to identify due to the weak electron scattering from the organic molecules, the CdS nanoparticles (cores) could be clearly observed. The nanoparticles were approximately spherical with no appreciable habit planes and some exhibited atomic columns because the crystallographic zone axis was approximately parallel to the incident electron direction. These results indicate that the average size of the CdS nanoparticles was 3.4 nm and twin boundaries were formed.

Figure 3 shows the HAADF-STEM images of the CdS-Dend nanoparticles without and with annealing. In the unannealed specimen, the nanoparticles looked agglomerated because of their inhomogeneous dispersion, as indicated by the red dotted lines and more clearly visible in the enlarged view shown in Fig. 3(c). By assuming that these were spherical agglomerations embedded in the thin foil specimen, we estimated an average size of about 16 nm. These results could suggest a change in the packing density of the nanoparticles in those regions. However, the electron tomography was hard to perform due to the significant radiation damage induced by a long-time electron exposure; thus, neither the three-dimensional structure nor the packing density of the agglomerations could be determined, so that understanding the detailed structure remains a challenge. Figures 3(b) and 3(d) show the HAADF-STEM images of a thin foil subjected to annealing. In contrast with the unannealed specimen, the nanoparticles were more uniformly distributed and no significant agglomeration was observed. These results demonstrate that the arrangement of the CdS-Dend nanoparticles could be changed by annealing. However, the $P2_13$ space group generated by annealing was difficult to identify.

To investigate the structural order, we used a two-dimensional autocorrelation function $A(x,y)$ [12–16], defined as follows [17]:

$$A(x,y) = \sum_{x_1, y_1} \frac{1}{N} I(x_1, y_1) I(x + x_1, y + y_1), \qquad (1)$$

where $I(x,y)$ is the intensity of the HAADF-STEM image at a pixel position $(x,y)$, $N$ is the number of sampling points in the summation, and $x_1$ and $y_1$ represent arbitrary pixel positions; the summation is performed over all the $x_1$ and $y_1$ values. The interparticle spacing of the nanoparticles can be deduced from the peak positions in the autocorrelation function.

Figure 4(a) displays the HAADF-STEM image of the unannealed specimen shown in Fig. 3(a) and the corresponding autocorrelation plot (inset), with two rings around the central spot. The presence of the rings indicates a short-range structural order even in the unannealed specimen. Figure 4(b) presents the corresponding results for the region of the annealed specimen shown in Fig. 3(b), still with two rings in the autocorrelation plot. The radial intensity distributions of these two autocorrelation patterns are plotted in Fig. 4(c); to improve the signal-to-noise ratio, the plots were averaged over circles of equal radius $r = \sqrt{x^2 + y^2}$. The peak positions determined for the unannealed specimen were 5.65 and 11.3 nm and those for the annealed one were 6.07 and 11.5 nm; hence, the first peak position of the annealed specimen is larger than that of the unannealed one, even if taking into account the measurement error (± 0.1 nm).

To examine the self-assembled structure in the annealed specimen in greater detail, we simulated the radial intensity profile of the autocorrelation function by taking into account the $P2_13$ unit cell revealed by the small-angle X-ray scattering measurements [11]. Figure 4(d) shows the two-dimensional supercell, consisting of 5 × 5 unit cells, used in the simulation; the red square indicates a unit cell and the yellow dots represent 4 nanoparticles in it, viewed along the [001] direction. To better approximate the STEM observations, this simulated image was subjected to Gaussian noise. The dashed plot in Fig. 4(c) represents the results of the simulation. The positions of the first and second simulated peaks (6.45 and 11.8 nm, respectively) were in reasonable agreement with the results for the annealed specimen (6.07 and 11.5 nm, respectively). This confirmed that the observation results (Fig. 3) were consistent with the $P2_13$ structural model.

The autocorrelation alone cannot determine the space group because it can only deduce the interparticle spacing from an electron microscope image. The space group and crystal symmetry in the self-assembled state can be discussed more straightforwardly based on the small-angle electron diffraction technique. However, due to the short-range order resulting from the self-assembly attained in the thin foil specimen, the Bragg reflections were too weak to be recognized. Although electron tomography can be another useful way for the symmetry analysis, the potential radiation damages did not allow the long-time exposure required to collect a tilt series of images. Interestingly, the autocorrelation plot for the annealed state (Fig. 4) showed well-defined peaks, with distinct positions from those of the unannealed state. In addition, the peak positions for the annealed state were not in contradiction with those of the $P2_13$ structure, which should be produced by such annealing [11]. Our electron microscopy observations accordingly provided useful information for the phase identification.

There were some discrepancies between simulation and experimental results, as shown in Fig. 4(c); the third simulated peak predicted was weak in the observation results and, in addition, the simulated positions of the first two peaks (6.45 and 11.8 nm) slightly differed from the observed ones (6.07 and 11.5 nm). These discrepancies were probably due to the simplified structural model used for the simulation (Fig. 4(d)), which only assumed the [001] projection although the actual specimens contained many ordered regions (crystal grains) with various orientations (i.e., polycrystals of the self-assembled structure). Nevertheless, according to the observations, the self-assembly induced by annealing shifted the first peak position of the ordered state toward a higher value compared to the disordered state, i.e., 6.07 nm for the annealed specimen and 5.65 nm for the unannealed specimen. As a result, the autocorrelation function of the annealed specimen gave peak positions closer to the simulated ones and, accordingly, confirmed that the self-assembly was promoted by the annealing treatment.

By assuming that a foil with an ordered phase is polycrystalline, the crystal grain size of the self-assembled structure can be a measure of the characteristic dimensions of the ordered regions. Hence, to determine this crystal grain size, some image processing steps were carried out as follows: (1) obtaining a digital diffractogram from the HAADF-STEM image shown in Fig. 3(b), (2) selection of a specific frequency region representing a Bragg reflection, and (3) image reconstruction using the frequency region selected. However, the contrast in the reconstructed image (not reported in this paper) was not sufficient to identify crystal grains. Thus, we could reasonably assume that the grain size was smaller than the foil thickness (90 nm). Since the original HAADF-STEM image is a projection, small crystal grains can be obscured due to some undesired superposition along the electron incidence direction. The results indicate that the $P2_13$ structural order can be achieved only in a range of tens of nanometers.

**Conclusions**

The STEM results obtained in this study on CdS-Dend nanoparticles led to the following conclusions:

(1) Ultramicrotomy is an effective and simple approach for the detailed structural analysis of self-assembled nanoparticles.

(2) The presence of the $P2_13$ structure after annealing could be identified by applying an autocorrelation function to the HAADF-STEM images of thin foil specimens.

(3) This structural order appears to be limited to a range of tens of nanometers since the crystal grains in the ordered phase could not be identified in the 90 nm thick foil specimens.

**Funding**

This study was supported in part by JST, CREST (JPMJCR1664) and JSPS KAKENHI (JP18H03845 and JP16H04190).

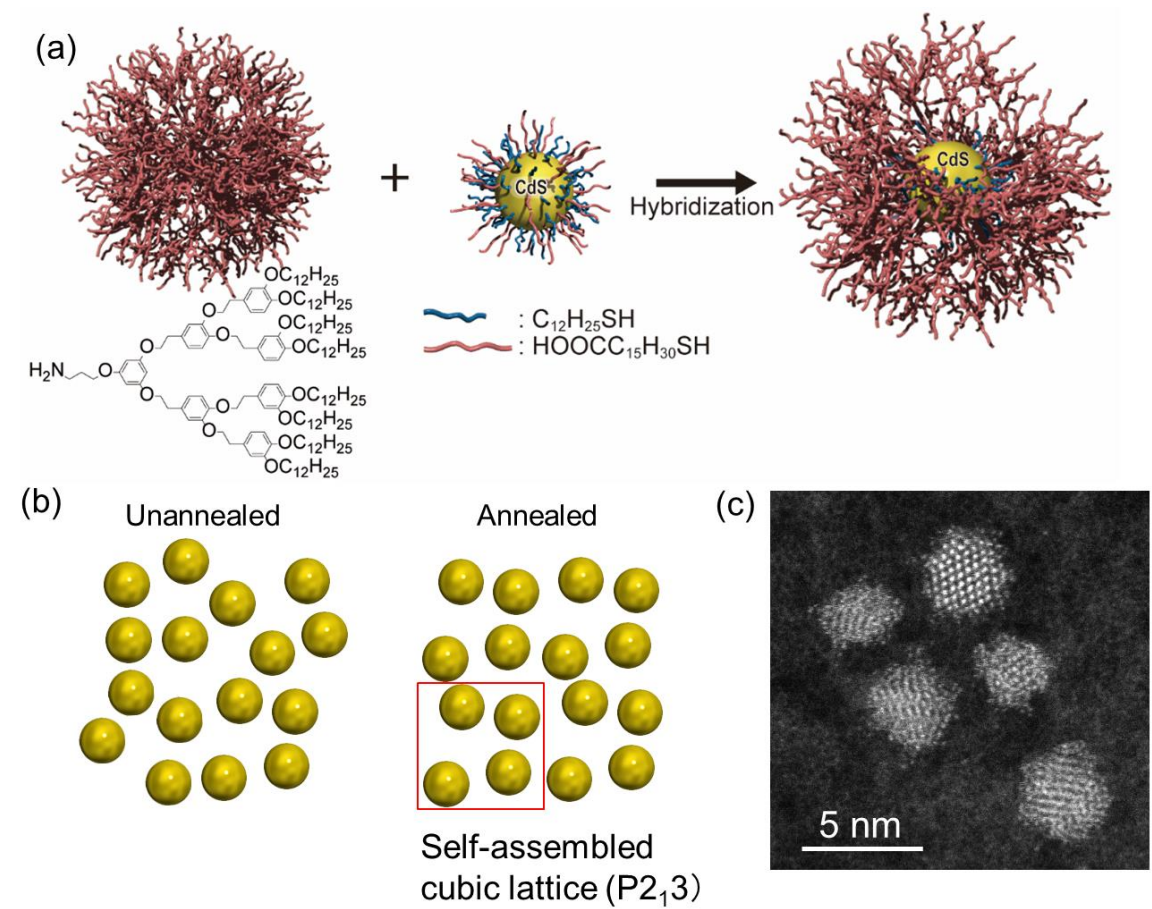

Fig. 1. (a) Schematic of (left) aromatic–aliphatic dendrons, (middle) a CdS nanoparticle modified by alkyl and carboxyl thiols, and (right) a CdS nanoparticle modified by these thiols and dendrons (dendronized CdS, or CdS-Dend). Figure reproduced with permission from Ref. [11]. (b) Schematic of the CdS-Dend structures without and with annealing; the annealing treatment converted the disordered structure into a self-assembled one with a cubic lattice and a $P2_13$ symmetry (The red square indicates a unit cell). (c) High-angle annular dark-field scanning transmission electron microscopy image of CdS-Dend nanoparticles dispersed on a carbon grid.

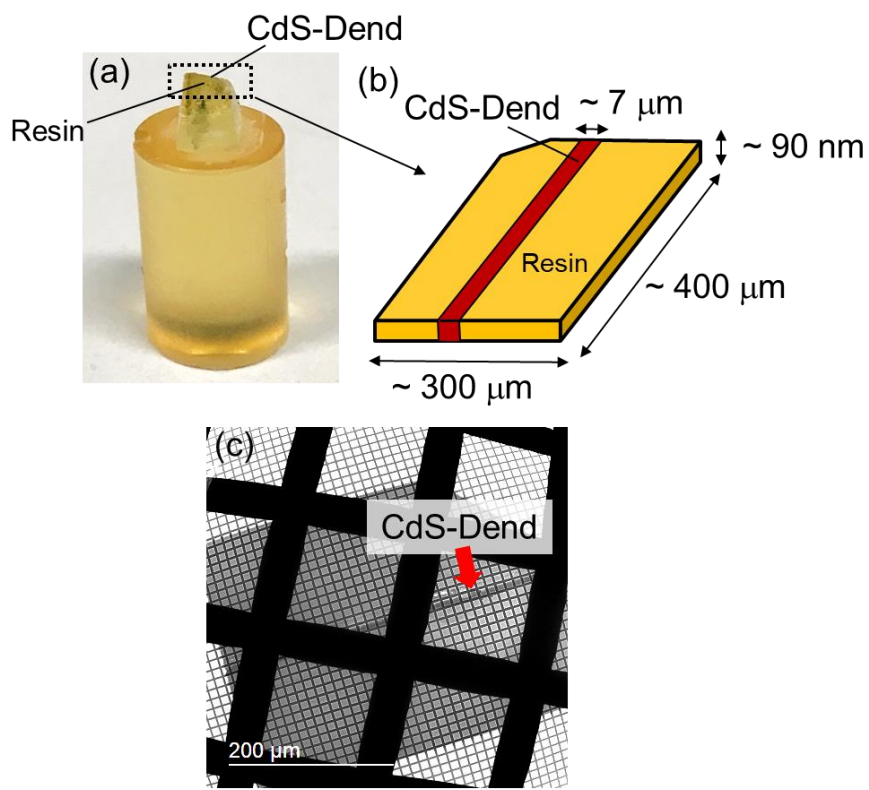

Fig. 2. (a) Dendronized CdS (CdS-Dend) nanoparticles embedded in resin for ultramicrotomy. (b) Schematic of a specimen cut from the rectangular area of the resin (dashed box) indicated in (a), containing the embedded CdS-Dend nanoparticles. (c) Bright-field transmission electron microscope image of a thin film, mounted on a carbon grid, prepared by ultramicrotomy. The large (~100 μm) and small (~7 μm) squares represent the copper and carbon meshes, respectively.

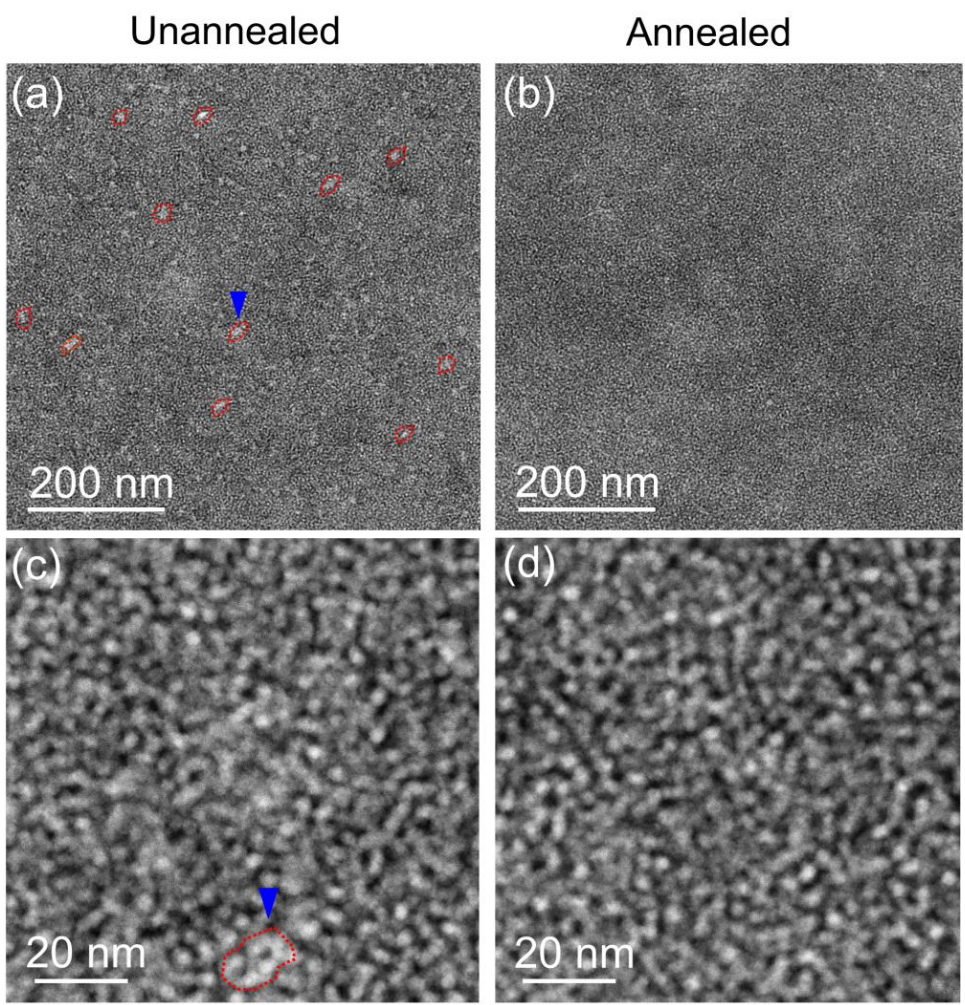

Fig. 3. Changes in dendronized CdS (CdS-Dend) nanoparticles after annealing. High-angle annular dark-field scanning transmission electron microscopy images of (a,c) unannealed and (b,d) annealed specimens. In (a), the CdS-Dend agglomerations are outlined with red broken lines and one of them (indicated by a blue arrowhead) is magnified in (c).

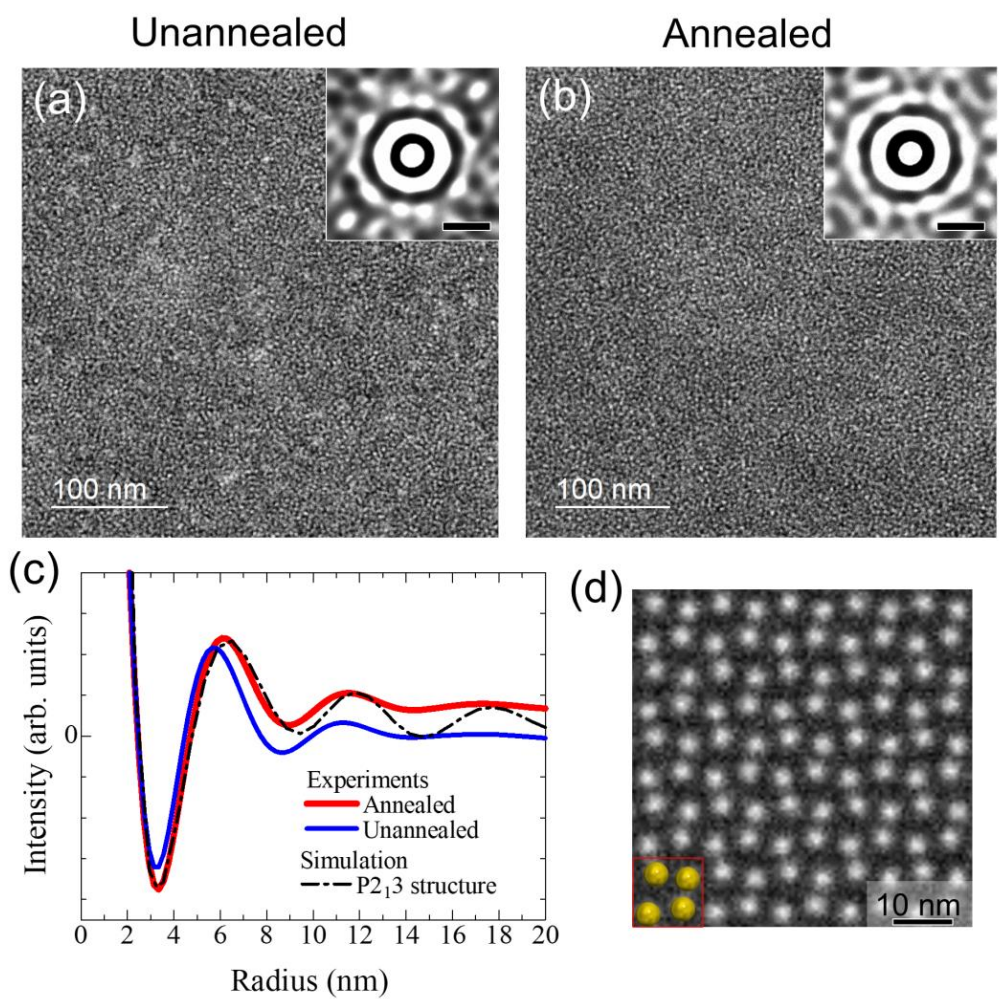

Fig. 4. Autocorrelation function analysis of dendronized CdS (CdS-Dend) nanoparticles. High-angle annular dark-field scanning transmission electron microscopy (HAADF-STEM) images of the (a) unannealed and (b) annealed specimens with the corresponding autocorrelation function plots in the insets. The scale bar in the insets is 10 nm. (c) Autocorrelation function against radius; the functions were calculated by averaging over all the in-plane directions for the insets in (a) and (b). (d) Simulated HAADF-STEM image of CdS-Dend nanoparticles arranged with the $P2_13$ structure, which was used to calculate the autocorrelation function. A schematic of the self-assembled CdS-Dend is displayed at the bottom left and the red square represents the unit cell of the $P2_13$ structure.

13